
\magnification=\magstep1

\hsize=6.5truein
\vsize=8.9 truein
\baselineskip=\normalbaselineskip \multiply\baselineskip by
2
\def\beginlinemode{\endmode\begingroup\parskip=0pt
\obeylines\def\\{\par}\def\endmode{\par\endgroup}}

\def\beginparmode{\endmode\begingroup
\def\endmode{\par\endgroup}}

\def\endpage{\vfill\eject}

\def\raggedcenter{\leftskip=4em plus 12 em
\rightskip=\leftskip \parindent=0pt \parfillskip=0pt
\spaceskip=.3333em \xspaceskip=.5em \pretolerance =9999
\tolerance=9999 \hyphenpenalty=9999 \exhyphenpenalty=9999 }

\let\endmode=\par{\obeylines\gdef\
{}}

\overfullrule=0pt
\medskipamount=7.2pt plus2.4pt minus2.4pt

\parskip=\medskipamount

\def\refto#1{$^{#1}$}

\gdef\refis#1{\indent\hbox to 0pt{\hss#1.~}}

\gdef\journal#1, #2, #3, 1#4#5#6{{\sl #1~}{\bf #2}, #3, (1#4#5#6)}

\def\prb{\journal Phys. Rev. B, }

\def\etal{{\it et al.}}

\def\prl{\journal Phys. Rev. Lett., }

\def\pr{\journal Phys. Rev., }

\def\rmp{\journal Rev. Mod. Phys., }

\def\bisco{$\rm Bi_2Sr_2CaCu_2O_8$}
\def\biscos{$\rm Bi_2Sr_2CaCu_2O_8$\ }
\def\bbisco{$\bf Bi_2Sr_2CaCu_2O_8$}
\def\bbiscos{$\bf Bi_2Sr_2CaCu_2O_8$\ }
\def\nbsn{$\rm Nb_3Sn$}
\def\nbsns{$\rm Nb_3Sn$\ }
\def\lasr{$\rm La_{1.85}Sr_{.15}CuO_4$}
\def\lasrx{$\rm La_{2-x}Sr_{x}CuO_4$}
\def\lasrs{$\rm La_{1.85}Sr_{.15}CuO_4$\ }
\def\ybco{$\rm YBa_2Cu_3O_7$}
\def\ybcos{$\rm YBa_2Cu_3O_7$\ }

\rightline{NSF-ITP-93-16}
\rightline{cond-mat/9301034}
\null\vskip 3pt plus 0.2fill \beginlinemode
\raggedcenter {\bf Lower Bound on the Energy Gap at Precise Locations
on the \bbisco Fermi Surface}

\vskip 3pt plus 0.2fill Michael E. Flatt\'e

\vskip 3pt plus 0.2fill {\sl Institute for Theoretical Physics\\
University of California\\Santa Barbara, CA  93106--4030}

\vskip 3pt plus 0.3fill \beginparmode \narrower Recent
inelastic-neutron-scattering measurements of a
low-energy phonon in \biscos provide enough information to place a lower
bound on the energy gap magnitude at precise locations on the Fermi surface.
Measurements of a few low-energy phonons should indicate whether the
gap symmetry is $d_{x^2-y^2}$. Further measurements could map out the
angle-resolved gap magnitude on the entire Fermi surface.


\footline={\hss\tenrm\folio\hss}

\endpage\beginparmode

Recently H.A. Mook, \etal\refto{1}, have examined the
temperature-dependence of phonon linewidths
in \bisco. The authors are primarily concerned with the behavior
of high-energy phonons ($\hbar\omega>50$meV) near and below $T_c$.  Their
results indicate strong anharmonicity and electron-phonon coupling for these
modes.
The purpose of this Communication is to point out that
the temperature-dependence
of the $5$meV phonon,
which was used as a control in this experiment, provides some
information about the superconducting energy gap at precise locations on
the Fermi surface.
In previous work\refto{2} I presented
a method, using measurements of phonon anomalies, to determine the
energy-gap anisotropy in a superconductor with a quasi-two-dimensional
Fermi surface. In this Communication I
will apply this method to the $5$meV phonon.
If the temperature-dependence of other low-energy
phonons were probed, more information about the anisotropy of the gap
could be obtained.

The behavior of phonon anomalies in superconductors without
quasi-two-dimensional Fermi surfaces is well-known.
If the superconducting gap is isotropic,
the minimum energy for a phonon to decay into two
quasiparticles is twice that gap. This produces an anomaly in a phonon
dispersion  curve at low temperature when the curve
crosses $\hbar\omega = 2\Delta$, where
$\Delta$ is the
superconductor's gap.  The anomaly also manifests itself in a
temperature-shift of the phonon's linewidth
across $T_c$ (observed first\refto{3} in $\rm Nb_3Sn$).
When the energy
gap is anisotropic, the anomaly is smoothed.

If the Fermi
surface is quasi-two-dimensional, however, the anomaly remains sharp
and is momentum-dependent in a useful way\refto{2}.
The minimum energy for a
phonon of wavevector $\vec q$
to decay into two quasiparticles in such a superconductor
depends on the gap
magnitude at a discrete number of points on the Fermi surface
(different for each
phonon).  These points can be determined by placing the phonon momentum
vector on the Fermi sea such that its head and tail are located on the Fermi
surface.  The gap magnitude at the location of the head and the location
opposite the tail determine the minimum energy according to the
following equation:
$$\tilde\Delta(\vec q) \equiv |\Delta(\vec k_1)|+|\Delta(\vec
k_2)|,\eqno(1)$$
where $\vec k_1$ and $\vec k_2$ are the two locations on the Fermi
surface described above and $\Delta(\vec k)$ is the superconductor's gap as
a function of momentum.  The locations $\vec k_1$ and $\vec k_2$ are the
{\it only} places where momentum conservation ($\vec k_1 +\vec k_2 =
\vec q$) can be satisfied {\it and} low-energy excitations exist.

The minimum, or threshold energy
$\tilde\Delta(\vec q)$ defines a surface in energy-momentum space.
The linewidths of phonons whose energies lie far above this surface
should be unaffected by a transition into the superconducting state.
The linewidths of phonons whose energies lie close to but above
the threshold surface should increase, and those below should decrease.
Information about the threshold surface can be directly inverted to
obtain the gap magnitude as a function of location on the Fermi
surface.
Thus by observing temperature shifts in many different
phonons, the gap magnitude can be mapped out everywhere on the Fermi
surface.

This method has been applied\refto{2}
to recent data\refto{4} on $\rm YBa_2Cu_3O_7$.  Currently a
group at Brookhaven\refto{5}
 is testing the feasibility of this theory on $\rm
La_{1.85}Sr_{.15}CuO_4$. Preliminary results are encouraging.

The observation of the $5$meV phonon changing linewidth in $\rm
Bi_2Sr_2CaCu_2O_8$ restricts the possible gap values at a few
isolated points on the Fermi surface.  There are three possible ways to
place the phonon momentum, [$\vec q=0.5(\pi,\pi)$]
 on the Fermi surface (taken from photoemission data\refto{6})
satisfying Equation (1).
Vectors $1$ and $2$ connect points on the same piece of Fermi surface (see
Figure),
while vector $3$ connects points on different pieces.  Assuming the gap
magnitude has the symmetry of the square lattice, these are the only
non-degenerate ways for this phonon to decay into electronic
excitations.
Since the phonon
is perpendicular to a line of reflection symmetry for the
quasi-two-dimensional Fermi surface, $|\Delta(\vec k_1)|=|\Delta(\vec
k_2)|$ for vectors $1$ and $2$.  This places a lower bound of $2.5$meV on
the gap magnitude at the locations marked with filled squares
and labeled $1$
and $2$. The Fermi surface locations probed by the vectors have been
mapped by lattice symmetry to a $45^o$ region in the first quadrant.
 The two squares labeled $3$ correspond to vector $3$,
which connects two locations not identical under a lattice
symmetry operation. The sum of the gap magnitudes at these points must
exceed $5$meV.
The observation that $5$meV is less than twice the gap at these
locations agrees with photoemission experiments\refto{7}.

Even though three different electronic decay channels exist for this
phonon, lower bounds can be set because points $1-3$ are very close to
each other.  It is conceivable, though unlikely, that the channel
associated with vector $1$ or $2$ dominates the other two. Then the
lower bound would only apply to point $1$ or $2$.

This complication would not affect observations of temperature
shifts in some other low-energy phonons in $\rm
Bi_2Sr_2CaCu_2O_8$.  Of particular interest would be any low-energy
phonons with momenta near $0.88(\pi,0)$. This momentum
is shown on the Figure, and
marked $\vec q_n$.  This is the only way to place the vector on the
Fermi surface such that Equation (1) is satisfied.
Photoemission indicates that the gap magnitudes
near the head and tail of $\vec
q_n$
are
substantially smaller than elsewhere on the Fermi surface\refto{7}.
This
experiment's resolution, however, is only about $10$meV.  Better
resolution should be available with phonons.

I would like to express thanks to D.S. Dessau and Z.-X. Shen for
permission to use their Fermi surface data prior to publication.  This
work was supported by the NSF under Grant No. PHY89-04035.

\vfill
\eject

\filbreak\vskip 0.5truein{\raggedcenter REFERENCES
\par}\nobreak\vskip 0.25truein\nobreak

\refis{1.} H.A. Mook, \etal, \prl 69, 2272, 1992.

\refis{2.} M.E. Flatt\'e, \prl 70, 658, 1993.

\refis{3.} J.D. Axe, G. Shirane, \prl 30, 214, 1973.

\refis{4.} N. Pyka, \etal, to be published.

\refis{5.} H. Chou, G. Shirane, M. Matsuda, R.J. Birgeneau, to
be published.

\refis{6.} D.S. Dessau, Z.-X. Shen, to be published.

\refis{7.} B.O. Wells, \etal, \prb 46, 11830, 1992.
Z.-X. Shen, \etal, unpublished.

\vfill\eject\vskip 0.5truein{\raggedcenter FIGURE CAPTION
\par}\nobreak\vskip 0.25truein\nobreak
\beginparmode

VectorS $1-3$ are the non-degenerate ways to position the phonon
momentum $\vec q=0.5(\pi,\pi)$ on the \biscos Fermi surface (dashed
line)\refto{6}. Since the $5$meV phonon narrows upon transition to
the superconducting state, the gap magnitudes at locations $1$ and $2$
(filled squares) must exceed $2.5$meV. The sum of the gap magnitudes at
points $3$ exceeds $5$meV. Observations of linewidth changes in phonons
with momenta near $\vec q_n$ would probe proposed nodes in the gap
magnitude\refto{7}.

\end